\documentclass[12pt]{article}
\usepackage{amsmath,amssymb}

\usepackage{graphicx}
\usepackage{wrapfig}
\usepackage{color}

\usepackage{hyperref}
\usepackage[sort,compress]{cite}

\def\beq{\begin{eqnarray}}
\def\eeq{\end{eqnarray}}

\def\lb{\label}

\newcommand{\fract}[2]{{\textstyle\frac{#1}{#2}}}

\newcommand{\be}{\begin{equation}}
\newcommand{\ee}{\end{equation}}
\newcommand{\bea}{\begin{eqnarray}}
\newcommand{\eea}{\end{eqnarray}}
\newcommand{\bg}{\begin{gather}}

\newcommand{\bseq}{\begin{subequations}}
\newcommand{\eseq}{\end{subequations}}

\def\be{\begin{eqnarray}}
\def\ee{\end{eqnarray}}
\def\lb{\label}

\begin{document}

\title{\textbf{\Large
Beltrami fields, dispersive electromagnetic waves and gravitational spheromaks  from chiral anomaly
}}
\vspace{2cm}
\author{ \textbf{  Sergey N. Solodukhin}} 

\date{}
\vspace{2cm}
\maketitle
\begin{center}
\emph{Institut Denis Poisson, 
UMR CNRS 7013, Universit\'e de Tours, \\
Parc de Grandmont, 37200 Tours, France}

\end{center}




\vspace{0.2mm}

\begin{abstract}

\noindent { In this note, we focus on the backreaction effects due to the chiral anomaly. The chiral anomaly modifies conserved currents, introducing new contributions. For Maxwell gauge fields, this leads to a contribution to the electric current proportional to the background magnetic field, a phenomenon known as the chiral magnetic effect, which is widely discussed in the literature. In the case of gravitational fields, as we demonstrate, the anomaly induces a new contribution to the stress-energy tensor.
We analyze the potential manifestations of these modifications in the gravitational field and hydrodynamics in chiral media, and we also comment on backreaction effects in electrodynamics. In each case, we observe the systematic appearance of Beltrami-type fields. In electrodynamics and hydrodynamics, a Beltrami field (e.g., magnetic field or fluid velocity) is a vector parallel to its own curl. We propose a generalization of Beltrami fields for tensorial gravitational perturbations, calling the respective solutions to the gravitational equations {\it gravitational spheromaks} by analogy with a similar phenomenon in electrodynamics. In the modified hydrodynamics of chiral media, vorticity asymptotically forms a Beltrami vector field in a generalized Gromeka-Beltrami flow.

}

\end{abstract}

\vskip 1 cm
\noindent
\rule{7.7 cm}{.5 pt}\\
\noindent 
\noindent

\noindent ~~~ {\footnotesize e-mail: sergey.solodukhin@univ-tours.fr}

\pagebreak

\section{Introduction}

Quantum anomalies play an increasingly important role not only in high-energy physics, where they were originally discovered, but also in condensed matter physics, where they may manifest in various new materials, such as Dirac and Weyl semimetals. Recent developments in condensed matter offer a rich playground for testing concepts from modern high-energy physics, demonstrating that the interplay between these two domains is mutually beneficial. For a recent review, see \cite{Chernodub:2021nff}. The focus of the present note is on quantum chiral anomalies \cite{Adler:1969gk}, \cite{Alvarez-Gaume:1983ihn}, which have found numerous applications in the physics of quantum fluids. For a far from exhaustive list of recent publications on the subject, see \cite{Alekseev:1998ds}-\cite{Stone:2018zel}.

In four dimensions, chiral symmetry is present at the classical level for massless fermions, described by the transformation
$\psi\rightarrow e^{i\alpha \gamma_5}\psi$. 
This symmetry manifests in the conservation of the corresponding chiral current, given by 
$\nabla_\mu j^\mu_5=0$, where $j_5^\mu=\bar{\psi}\gamma^\mu \gamma_5\psi$.
However, after quantization, the conservation of the chiral current is no longer valid due to the so-called chiral anomaly. 
This anomaly arises when fermions couple to external fields, such as a gauge field
gauge field $A_\mu$ or a gravitational field described by the metric $g_{\mu\nu}$. 
Consequently, the conservation law is modified by anomaly terms, as shown in \cite{Adler:1969gk},  \cite{Alvarez-Gaume:1983ihn}:
\be
\nabla_\mu j^\mu_5= -\frac{\sigma_F}{16\pi^2}\epsilon^{\alpha\beta\mu\nu}F_{\alpha\beta}F_{\mu\nu}-\frac{\sigma_R}{384\pi^2}\epsilon^{\alpha\beta\mu\nu}R_{\alpha\beta\sigma\rho}
R^{\sigma\rho}_{\ \ \mu\nu}\, \, ,
\lb{1}
\ee
where $F_{\mu\nu}=\partial_\mu A_\nu-\partial_\nu A_\mu$ and $R_{\alpha\beta\mu\nu}$ is the Riemann tensor defined for the metric $g_{\mu\nu}$. Here $\sigma_F$ and $\sigma_R$ are the respective chiral charges.
While the generalization to non-abelian gauge fields is straightforward, this note will focus on the abelian case for simplicity.

Recently, several possible manifestations of the chiral anomaly, both due to Abelian gauge fields and gravitational effects, have been discussed in the context of chiral media. The most popular and perhaps the most intriguing is the Chiral Magnetic Effect (CME) \cite{Alekseev:1998ds}, \cite{Fukushima:2008xe}, \cite{Kharzeev:2013ffa}, which predicts that, in the presence of a chiral anomaly, an electric current proportional to the background magnetic field arises: $\vec{j}=\kappa \vec{B}$.
In this note, we provide a simple derivation of this relation and further generalize it to the gravitational case. In the latter, we compute the modifications to the stress-energy tensor due to the anomaly and explore several potential manifestations of these modifications. Some gravitational aspects of the chiral anomaly in applications to quantum fluids were previously considered in \cite{Prokhorov:2022udo}.

While electromagnetic effects in chiral media have been discussed in the literature (though certain aspects, such as the rotation of the polarization plane and negative refraction \cite{ngt}, are not included in this note), the discussion of gravitational effects, which is the primary focus here, is entirely original.

\section{Maxwell  field chiral anomaly}

\medskip

\noindent {\it Anomaly action.} It is convenient to promote the chiral transformations  to a local symmetry by defying $\psi\rightarrow e^{i\alpha(x) \gamma_5}\psi$  and introducing a  respective gauge field $A_5^\mu$ 
whose  transformation compensates for the gradients of $\alpha(x)$,  $A^5_\mu\rightarrow A^5_\mu -\partial_\mu \alpha(x)$.
Provided that $W$ is the respective effective action the chiral current is defined as $j_5^\mu=\frac{\delta W}{\delta A^5_\mu}$ and the anomaly then manifests as non-vanishing divergence 
$\nabla_\mu \frac{\delta W}{\delta A^5_\mu}$.
The effective action, that generates the chiral anomaly (\ref{1}), can be easily  constructed if considered as a functional of both $A^5_\mu$ and $A_\mu$.
Indeed, noting that $\epsilon^{\alpha\beta\mu\nu}F_{\alpha\beta}F_{\mu\nu}=\nabla_\mu S^\mu$, $S^\mu=2\epsilon^{\mu\nu\alpha\beta} A_\nu F_{\alpha\beta}$
one finds that the anomaly part in the effective action takes the form
\be
W_{A}[A_\mu^5, A_\mu]=-\frac{\sigma_F}{32\pi^2}\int d^4x \sqrt{-g}A^5_{\mu}S^\mu
\, .
\lb{2}
\ee

\bigskip

\noindent {\it Chiral magnetic effect.}  The electric current is defined as variation,  $j_\mu=\frac{\delta W}{\delta A^\mu}$, of the action with respect to the abelian gauge field $A_\mu$.
Thus, the current due to anomaly is 
\be
j^\mu=\frac{\sigma_F}{16\pi^2} (2\epsilon^{\mu\nu\alpha\beta}A^5_\nu F_{\alpha\beta}-\epsilon^{\mu\nu\alpha\beta}A_\nu {\cal F}_{\alpha\beta})\, ,
\lb{3}
\ee
where ${\cal F}_{\alpha\beta}=\partial_\alpha A^5_\beta-\partial_\beta A^5_\alpha$. The chiral gauge field is often considered to be a constant vector with only one non-vanishing
component $A^5_0=\mu_5$ associated with the chiral chemical potential $\mu_5=\frac{1}{2}(\mu_L-\mu_R)$. In this case one may disregard the last term in (\ref{3}).  The electric current derived from (\ref{3})  is
\be
j_i=\kappa B_i\, ,  \  \  \  \  \kappa=\frac{\sigma_F \mu_5}{8\pi^2}\, 
\lb{4}
\ee
where $B_i=\epsilon_{ijk}F_{jk}$ is the magnetic field, that is precisely the chiral magnetic effect.

\bigskip

\noindent {\it Maxwell equations with backreaction.} We now look at the backreaction of the electric current (\ref{4}) on the Maxwell field and, in particular, on what modifications it causes for the
propagation of the electro-magnetic waves. In the absence of any other currents and electric charges the Maxwell equations take the form
\be
&&\partial_i E_i=\partial_i B_i=0\, , \  \  \   \nonumber \\
&&\epsilon_{ijk}\partial_j E_k+\partial_t B_i=0 \,  ,\  \  \  \nonumber \\
&&\epsilon_{ijk}\partial_j B_k-\partial_t E_i=\kappa B_i\, ,
\lb{5}
\ee
where $t\equiv  x^0$ and the electric and magnetic fields are defined as usual $E_i=F_{0i}$ and $B_i=\epsilon_{ijk}F_{jk}$.
In what follows we assume that $\kappa>0$.

\bigskip

\noindent {\it Electromagnetic waves.} Following the standard 
electromagnetic wave analysis \cite{Jackson} we consider a plane wave $E_i=E_0\epsilon^{(1)}_{i} e^{-\imath\omega t+\imath p_i x^i}$ and $B_i = B_0 \epsilon^{(2)}_{i} e^{-\imath \omega t+
\imath p_i x^i}$.
Here constants $E_0$ and $B_0$ represent the amplitude of  respectively electric and magnetic fields and $\epsilon^{(1)}_i$ and $\epsilon^{(2)}_i$ are two unitary complex vectors. Substituting this plane wave into (\ref{5})
we find that the two unitary vectors are orthogonal to the direction of the pane wave, $p_i \epsilon^{(1)}_i=p_i \epsilon^{(2)}_i$.  The second line equation in (\ref{5})
tells us that there is a relation between the two unitary vectors,
\be
\epsilon^{(2)}_i=\frac{E_0}{\omega B_0}\epsilon_{ijk} p_j \epsilon^{(1)}_k\, ,
\lb{6}
\ee
so that the three vectors $\vec{p}, \  \vec{\epsilon}^{(1)}, \  \vec{\epsilon}^{(2)}$ are mutually orthogonal.  The normalization condition for the unitary vectors then gives a relation,
$|B_0/E_0|=p/\omega$.

If $\kappa=0$ then the third equation in (\ref{5}) gives a similar relation $\epsilon^{(1)}_i=-\frac{B_0}{\omega E_0}\epsilon_{ijk} p_j \epsilon^{(2)}_k$. Combining these two relations one finds 
 the relativistic dispersion relation
$\omega^2=p^2$, $p^2=p_ip_i$ and $B_0=E_0$.
This describes a  relativistic electromagnetic wave.
Two independent components of vector $\epsilon^{(1)}_i$   define  two polarizations of the electromagnetic wave. If axis $x^1$ is chosen parallel to vector $\vec{p}$ then
these two independent directions are along axis $x^2$ and axis $x^3$.

In the present case, however,  $\kappa\neq 0$ and the third line equation in (\ref{5}) gives a new constraint
\be
\frac{B_0}{E_0}\epsilon_{ijk}p_j \epsilon^{(2)}_k+\omega \epsilon^{(1)}_i=-\imath \kappa \,\epsilon^{(2)}_i\, .
\lb{7}
\ee
Using relation (\ref{6}) one finds
\be
(\omega^2-p^2)\epsilon^{(1)}_i=-\imath \kappa \epsilon_{ijk}p_j\epsilon_k^{(1)}\, .
\lb{8}
\ee
Squaring this relation one finds a new, non-relativistic,  dispersion relation $(\omega^2-p^2)^2=\kappa^2p^2$ that has two branches,
\be
\omega^2=p^2\pm \kappa p\, ,
\lb{9}
\ee
where $p=\sqrt{p^2}>0$.  
In the negative sign branch the frequency $\omega$ becomes imaginary for $p< \kappa$. This  was noted before \cite{Shovkovy:2021yyw}, \cite{Yamamoto:2023uzq}. It signals about some instability\footnote{One could also add the Ohm current $\vec{j}=\sigma \vec{E}$ with conductivity $\sigma$. The respective instable modes  then are still defined by condition $p<\kappa$
independently of $\sigma$. This is known as chiral plasma instability \cite{inst}. I thank N. Yamamoto for remarks on this point.}.
The dispersion relation (\ref{9}) implies that 
\be
|B_0/E_0|=\frac{p}{\sqrt{p^2\pm \kappa p}}\, .
\lb{9-1}
\ee
The magnetic and electric components in the electromagnetic wave  no more have equal amplitude. Depending on value of $p$ and  the chosen sign branch  the magnetic (or electric)
component can be significantly  amplified with respect to the other component.

\bigskip

\noindent {\it Polarizations.}
Let us choose the coordinate system in which axis $x^1$ is parallel to
vector $\vec{p}$.  Then, provided that dispersion relation (\ref{9}) is valid, equation (\ref{8}) gives a relation $\epsilon^{(1)}_{2}=\pm \imath \epsilon^{(1)}_{3}$. The other relations come from (\ref{6}): $\epsilon^{(2)}_2=-\epsilon^{(1)}_3$ and $\epsilon^{(2)}_3=\epsilon^{(1)}_2$. With the normalization condition one finds for the unitary vectors:
$\vec{\epsilon}^{\, (1)}=\frac{1}{\sqrt{2}}(0, \pm \imath, 1)$ and  $\vec{\epsilon}^{\, (2)}=\pm \imath \,  \vec{\epsilon}^{\, (1)}$. For two different choices of sign the respective vectors 
$\vec{\epsilon}^{\, (1)}_+$ and $\vec{\epsilon}^{\, (1)}_-$ can be recognized as a pair of polarization vectors, see Jackson's book \cite{Jackson}. One finds then that in each sign branch one has a single polarization:  
\be
\vec{E}=(E_+ \vec{\epsilon}^{\ (1)}_+ e^{-\imath \omega_+ t}+E_- \vec{\epsilon}^{\ (1)}_- e^{-\imath \omega_- t})e^{\imath p x_1}\, ,
\lb{9-2}
\ee
where $\omega_\pm$ is the respective solution of equation (\ref{9}). Thus, only modes with $(-)$ polarization are instable for $p<\kappa$.

\bigskip

\noindent {\it Superluminal dispersive  waves.} 
The background chiral gauge fields $A^5_0$ breaks the Lorentz symmetry.  So that it is not surprising that the dispersion relation (\ref{9}) is non-relativistic.  
It is interesting that the respective group velocity is different from the speed of light.
Indeed, defining the group velocity as $v(p)=\frac{\partial\omega(p)}{\partial p}$ one finds
\be
v(p)=\frac{p\pm \frac{\kappa}{2}}{\sqrt{p^2\pm \kappa p}}\, .
\lb{10}
\ee
Notice that in the domain where $\omega(p)$ is real one has that $v(p)>1$, i.e. the group velocity is always larger than the speed of light (note that in our normalization $c=1$ is the speed of light in the chiral medium).
In this  sense (group velocity exceeds speed of light in the medium) the electromagnetic wave in the presence of the chiral  anomaly (i.e. chiral magnetic effect) is superluminal.  

The medium with the non-vanishing chiral chemical potential thus represents a dispersive medium. Its index of refraction $n(p)=p/\omega(p)$ is 
\be
n(p)=\sqrt{\frac{p}{p\pm \kappa}}\, .
\lb{11}
\ee
For the positive sign branch, it is less than one, $n(p)<1$, while for the negative sign branch it is larger than one, $n(p)>1$.

\bigskip

\noindent{\it Chiral electromagnetic spheromak.}
The frequency $\omega(p)$ vanishes when $p^2=\kappa^2$. This is a static solution to the Maxwell equations (\ref{5}). It lies on the border of the stability region $p> \kappa$. 
When both electric and magnetic fields 
are not time dependent one finds that\footnote{We use definition $(\textup{curl} \, \vec{a})_i=\epsilon_{lki}\partial_l a_k$.}   
\be
\textup{curl}\, \vec{B}=\kappa\vec{B}\, .
\lb{B}
\ee
This  defines the so-called force-free magnetic field. Solution to this equation was discussed by Chandrassekhar and Kendall
\cite{Ch-K}.  Squaring the operators one finds that the magnetic field satisfies a vector Helmholtz equation
$(\Delta+\kappa^2)\vec{B}=0$.   In plasma the lines of the magnetic field that solve equation
(\ref{B}) form the so-called spheromak. 

In hydrodynamics, equations similar to (\ref{B}) generically with position dependent $\kappa$,  are  known to define the Beltrami vector field. The  Beltrami vector fields have a close connection to Lagrangian turbulence. In the context of fluid mechanics the chaotic nature of the flux generated by these fields is discussed in \cite{ABC}, \cite{Beltrami}, see also \cite{Fre:2022odf}. It is natural to expect that in the systems, such as the quark-gluon plasma and Dirac or Weyl semimetal, 
in which the chiral magnetic effect appears,  the spheromak phenomenon should also be present, see also \cite{Chernodub:2010ye} for a relevant discussion.
 For an earlier discussion of the Beltrami fields in the context of the chiral magnetic anomaly 
see \cite{Wiegmann:2022syo}. For a recent discussion of a similar phenomenon in superconductors see \cite{Garaud:2022uof}.

\section{Gravitational chiral anomaly}

\noindent{\it Anomaly action and stress-energy tensor.} For the gravitational term in the chiral anomaly (\ref{1}) one introduces a local Chern-Simons vector $K^\mu$ such that $\nabla_\mu K^\mu=\frac{1}{4} \epsilon^{\alpha\beta\mu\nu}R_{\mu\nu\sigma\rho}R_{\alpha\beta}^{\ \ \sigma\rho}$. It is expressed in terms of the Christoffel symbols $\Gamma^\mu_{\alpha\beta}(g)$, see \cite{Jackiw:2003pm},
\be
K^\mu=-2\epsilon^{\mu\alpha\beta\gamma}(\fract {1}{2}\Gamma^\sigma_{\alpha\tau}\partial_\beta \Gamma^\tau_{\gamma\sigma}+\frac{1}{3}\Gamma^\sigma_{\alpha\tau}\Gamma^\tau_{\beta\eta}\Gamma^\eta_{\gamma\sigma})\, .
\lb{11-1}
\ee
The quantum action that generates the gravitational part in the chiral anomaly then is represented as a functional of
the chiral gauge field $A^5_\mu$ and the metric $g_{\alpha\beta}$,
\be
W_{A}[A^5_\mu, \, g_{\alpha\beta}]=-\frac{\sigma_R}{96\pi^2} \int d^4x \sqrt{-g} A^5_\mu K^\mu\, .
\lb{12}
\ee
 The respective stress-energy tensor obtained as a metric variation of (\ref{12}) reads \cite{Jackiw:2003pm}, \cite{Grumiller:2008ie}
\be
T^{A}_{\mu\nu}= -\frac{\sigma_R}{96\pi^2}  C_{\mu\nu}\, , \  \ C^{\mu\nu}=A^5_\alpha \epsilon^{\alpha\beta\sigma(\mu}\nabla_\sigma R^{\nu )}_{\ \beta}+
\nabla_{(\alpha}A^5_{\beta)}\, {^\ast}R^{\beta(\mu\nu)\alpha}\, ,
\lb{13}
\ee
where ${^\ast}R^{\beta\mu\nu\alpha}=\frac{1}{2}\epsilon^{\beta\mu}_{\  \ \sigma\rho}R^{\sigma\rho\nu\alpha}$.
Tensor $C_{\mu\nu}$ is a four-dimensional generalization of the Cotton tensor which plays an important  role in  three-dimensional gravity. It is traceless, $C^{\mu\nu}g_{\mu\nu}=0$
and its divergence is $\nabla_\mu C^{\mu}_{\ \nu}=-\frac{1}{8}A^5_\nu \, {^\ast}RR$, see \cite{Jackiw:2003pm}.  We will be interested in a modification of the Einstein equations due to
the anomaly term (\ref{13}).  Since all other terms in the Einstein equations are divergence free this imposes a constraint that ${^\ast}RR=0$, as discussed in   \cite{Jackiw:2003pm}.
This constraint, however, is trivially fulfilled in the linear field approximation that we consider below.
In the presence of a classical matter with stress-energy tensor $T_{\mu\nu}$ the Einstein equations then take the form
\be
R_{\mu\nu}-\frac{1}{2}g_{\mu\nu}R=8\pi G(T_{\mu\nu}+T^A_{\mu\nu})\, .
\lb{13-1}
\ee 
In vacuum, when no classical matter is present, the modified Einstein equations take the form
\be
R_{\mu\nu}=8\pi G T^{A}_{\mu\nu}\, ,
\lb{14}
\ee
where it is taken into account that $T^A_{\mu\nu}$ is trace free.
In what follows we assume that only one component of the axial gauge field is non-vanishing that is identified with the chiral chemical potential, $A^5_0=\mu_5=const$.

\bigskip

\noindent{\it Chiral energy and momentum fluxes. }  
In what follows we consider a weak field approximation representing the metric as $g_{\mu\nu}=\eta_{\mu\nu}+h_{\mu\nu}$, where $\eta_{\mu\nu}$ is the Minkowski metric and
$h_{\mu\nu}$ is a small perturbation. Following \cite{SC} we will use the following parametrization of the perturbations: $h_{00}=-2\Phi$, $h_{0i}=w_i$, $h_{ij}=s_{ij}-2\delta_{ij}\Psi$, $s_{ij}\delta^{ij}=0$. One can use the coordinate freedom and impose the conditions: $\partial_i w_i$=0, $\partial_{j}s_{ij}=0$. So that $s_{ij}$ represents the transverse-traceless component of the metric.

 Assuming that the  only
non-vanishing component for the chiral gauge field is $A^5_0=\mu_5$ one finds that to the leading order in perturbation the second term in $C_{\mu\nu}$ (\ref{13}) does not contribute and
that 
\be
T^A_{\mu\nu}=2\Lambda \epsilon_{0\beta\sigma (\mu}\partial^\sigma R_{\nu )}^{\ \beta}\, , \  \  \   \Lambda=\frac{\sigma_R}{192 \pi^2}
\lb{14-1}
\ee
For components one finds
\be
&& T^A_{00}=0\, , \nonumber \\
&&T^A_{0i}=  \Lambda \epsilon_{kli}\partial_l R_{0k}\, ,       \nonumber \\
&&T^A_{ij}=\Lambda (\epsilon_{kli}\partial_l R_{kj}+\epsilon_{klj}\partial_l R_{ki})\, ,
\lb{14-2}
\ee
where we define $\epsilon_{kli}=\epsilon_{0kli}$. Using  the components of the Ricci tensor given in appendix A one finds for the non-vanishing components
of the  induced stress-energy tensor
\be
&&T^A_{0i}=-\frac{\Lambda}{2} \epsilon_{kli}\partial_l\Delta w_k\, , \nonumber \\
&&T^A_{ij}=-\frac{\Lambda}{2} [\epsilon_{kli}(\partial_0\partial_l\partial_{j}w_{k}+2\Box\partial_l s_{kj})+(i \leftrightarrow  j)]\, ,
\lb{14-3}
\ee
where $\Delta=\partial_i\partial_i$, $\Box=-\partial_0^2+\Delta$.
This demonstrates non-trivial momentum and stress fluxes while the energy density remains zero. The status of equations (\ref{14-1})-(\ref{14-2}) is analogous to that of the chiral magnetic relation (\ref{4}). In a chiral medium subjected to a gravitational field characterized by non-trivial metric components $h_{\mu\nu}$, non-trivial energy and momentum fluxes are predicted to arise due to the chiral anomaly. Clearly, in the presence of only a Newtonian gravitational potential (component $\Phi$), this phenomenon does not occur. Rotation ($w_i$) or strain $(s_{ij})$  in the background metric is required for this new effect to manifest.

\bigskip

\noindent{\it The modified Einstein equations. Gravitational spheromak. }  We will now discuss the complete Einstein equations (\ref{14})  and their possible solutions.
In the weak field approximation the Einstein equations  (\ref{14}) reduce to
\be
G_{\mu\nu}=2\lambda \epsilon_{0\beta\sigma (\mu}\partial^\sigma R_{\nu)}^{\ \beta}\, , \   \   \  \lambda=8\pi G\Lambda
\lb{15}
\ee
where $G_{\mu\nu}=R_{\mu\nu}-\frac{1}{2}g_{\mu\nu}R$. In components one has that
\be
&&G_{00}=0\, , \nonumber\\
&&G_{0i}={\lambda}\epsilon_{kli}\partial_l R_{0k}\, , \nonumber\\
&&G_{ij}={\lambda}(\epsilon_{kli}\partial_l R_{kj}+(i \leftrightarrow  j))\, .
\lb{a2}
\ee 
A simple analysis presented in appendix A shows that restricting ourselves to the bounded solutions one has that $\Phi=\Psi=0$ and the components $w_i$ and $s_{ij}$
satisfy the following equations
\be
&&w_i-\lambda \epsilon_{kli}\partial_l w_{k}=0\, , \nonumber \\
&&s_{ij}-{\lambda}(\epsilon_{kli}\partial_l s_{jk}+\epsilon_{klj}\partial_l  s_{ik})=s^{0}_{ij}\,  , \  \  \Box s^{0}_{ij}=0\, .
\lb{16-1}
\ee
Tensor $s^{0}_{ij}$ thus represents the usual propagating gravitational wave with two polarizations, see also discussion in \cite{Jackiw:2003pm}.
The wave serves as a source in the tensor equation given in 
 (\ref{16-1}).
Assuming that the gravitational wave is absent ($s^{0}_{ij}=0$)
one finds that  equations (\ref{16-1}) can be written in the form\footnote{Note that the right hand side of second equation in (\ref{17}) is already symmetric. Indeed, its antisymmetric part $\epsilon^{mij}\epsilon_{kli}\partial_l  s^{}_{jk}=\partial_m s^{}-\partial_k s^{}_{km}=0$ since $s_{ij}$ is traceless and divergence free.}
\be
&&w_i={\lambda}\epsilon_{kli}\partial_l w_k\, , \nonumber \\
&&s_{ij}=2{\lambda}\epsilon_{kli}\partial_l  s_{kj}\, .
\lb{17}
\ee
Note in passing that equations  (\ref{17})  are similar to those that appear in three-dimensional gravity with the gravitational Chern-Simons term, see \cite{3d}.
Squaring the differential operator in (\ref{17}) one finds  the Helmholtz type equations
\be
&&(\Delta  +\lambda^{-2})w_i=0\, , \nonumber \\
&&(\Delta +(2\lambda)^{-2}) s^{}_{ij}=0\, ,
\lb{18}
\ee
where $\Delta=\partial_i \partial _i$.   Tensor $s_{ij}$ could be considered as a set of three vectors with index $j$ labeling the vectors.
Then  equations (\ref{17}) describe effectively four Beltrami vector fields. One should not, however,  overlook  the symmetry conditions $s_{ij}=s_{ji}$ that imply extra constraints.
Consider first the vectorial equation in (\ref{17}). Assuming that $w_i$ varies only along the $x_1$ axis one may represent $w_i=H^+_i e^{\frac{ix_1}{\lambda}}+H^-_ie^{-\frac{ix_1}{\lambda}}$
and find the relation $H_1^\pm=0$ and $H^\pm_2=\pm i H^\pm _3$. The other condition to be imposed is the reality of $w_i$. So that one finds that $w_{1}=0$ and the components
$w_{2} $ and $w_{3}$ are certain linear combinations of $\sin \frac{x_1}{\lambda}$ and $\cos \frac{x_1}{\lambda}$. Similar analysis can be done for components changing along axis $x_2$ and along  axis $x_3$. The complete solution to equations of this type, as given in (\ref{17}), is a linear combination of all three such solutions. We thus come to the following solution,
\be
&&w_1=\gamma \sin\frac{x_2}{\lambda}+\epsilon \cos \frac{x_2}{\lambda}-\chi \sin\frac{x_3}{\lambda}-\eta\cos\frac{x_3}{\lambda} \, , \nonumber \\
&&w_2=-\alpha \sin\frac{x_1}{\lambda}-\beta \cos \frac{x_1}{\lambda}+\chi \cos\frac{x_3}{\lambda}-\eta\sin \frac{x_3}{\lambda} \, , \nonumber \\
&&w_3=\alpha \cos \frac{x_1}{\lambda}-\beta\sin\frac{x_1}{\lambda}+\gamma \cos \frac{x_2}{\lambda}-\epsilon\sin\frac{x_2}{\lambda}\, ,
\lb{18-2}
\ee
with six parameters: $\alpha, \, \beta, \, \gamma, \, \epsilon, \, \chi, \, \eta$.  
For $\lambda=-1$,  one may impose extra conditions $\gamma=\eta=\beta=0$   and re-name parameters: $A=\chi$, $B=\alpha$ and $C=\epsilon$. Equation (\ref{18-2}) then
describes  the ABC (Arnold-Beltrami-Childress)  vector fields  widely discussed in the literature on fluid mechanics.

The second (tensorial) equation in (\ref{17}) is formally equivalent to  four vectorial equations. So that adding extra index $j$ to parameters in (\ref{18-2}) one finds a solution
(with $6\times 3=18$ parameters)  to the tensorial equation. The symmetry conditions $s^{}_{12}=s^{}_{21}$, $s^{}_{13}=s^{}_{31}$ and $s^{}_{23}=s^{}_{32}$ are, however, quite restrictive and leave only six independent parameters. Hence, one finds
\be
&&s_{12}=C\cos\frac{x_3}{2\lambda}-D\sin\frac{x_3}{2\lambda}=s_{21}\, , \nonumber \\
&&s_{13}=-B\sin\frac{x_2}{2\lambda}+A\cos \frac{x_2}{2\lambda}=s_{31}\, , \nonumber \\
&&s_{23}= -E\sin\frac{x_1}{2\lambda}-F\cos \frac{x_1}{2\lambda}=s_{32}\, , \nonumber \\
&&s_{11}=A\sin \frac{x_2}{2\lambda}+ B \cos \frac{x_2}{2\lambda}-C \sin\frac{x_3}{2\lambda}-D \cos\frac{x_3}{2\lambda}\, , \nonumber \\
&&s_{22}=F\sin \frac{x_1}{2\lambda}-E \cos\frac{x_1}{2\lambda}+D\cos\frac{x_3}{2\lambda}+C\sin\frac{x_3}{2\lambda}\, , \nonumber \\
&&s_{33}=E\cos\frac{x_1}{2\lambda}-F\sin\frac{x_1}{2\lambda}-B\cos \frac{x_2}{2\lambda}-C\sin\frac{x_2}{2\lambda}\, .
\lb{18-3}
\ee
Note that parameters here could be in principle functions of time.

Equations (\ref{17})?(\ref{18-3}) describe the non-trivial configurations of the gravitational field that we refer to as {\it gravitational spheromaks}. These configurations can be characterized by the behavior of the geodesic motion of test particles.
Considering the static case and only the terms linear in 3-velocity $v^i, \ i=1,\, 2,\, 3$ one finds\footnote{The  tensorial components $h^{}_{ij}$ would appear in the terms quadratic in velocity $v^i$.} (see discussion in \cite{SC})
\be
\frac{d\vec{v}}{d\tau}=\vec{v}\times \textup{curl}\vec{w} 
                                 = -\frac{1}{\lambda} \vec{v} \times \vec{w}\, ,
\lb{18-4}
\ee
where $w_i=h^{}_{0i}$. Notice that this equation is similar to the equation  for a charged test particle under  the Lorentz force in electrodynamics. This makes the analogy with the electrodynamic spheromak quite complete.

\bigskip

\noindent{\it  Hydrodynamics.}
Let us consider the case when there presents some classical matter with stress-energy tensor $T_{\mu\nu}$ in (\ref{13-1}). Then in a perturbation theory with
$\mu_5$ being the perturbation parameter one may rewrite (\ref{13-1}) as  the Einstein equations over Minkowski spacetime with a modified stress tensor
\be
&&\hat{T}_{\mu\nu}=T_{\mu\nu}+T^A_{\mu\nu}\, , \nonumber  \\
&&T^A_{00}=0\, , \nonumber \\
&& T^A_{0i}={\lambda} \, \epsilon_{kli}\partial_l T_{k0}\, , \nonumber \\
&&T^A_{ij}={\lambda} (\epsilon_{kli}\partial_l T_{kj}+\epsilon_{klj}\partial_l T_{ki})\, ,
\lb{19}
\ee
where ${\lambda}=8\pi G\Lambda$ and $T^A_{\mu\nu}$ is expressed in terms of the components of $T_{\mu\nu}$. 
Notice that $T^A_{0i}$  is automatically divergence free, $\partial_i T^A_{0i}=0$.
We are in particular interested in the case when the classical matter is represented by
viscous liquid. Then one has that
\be
&&T_{\mu\nu}=(\rho+p)u_\mu u_\nu +p \eta_{\mu\nu}-\sigma_{\mu\nu}\, , \nonumber \\
&&\sigma_{00}=0\, , \  \ \sigma_{0i}=0\, , \  \  \sigma_{ij}=\eta (\partial_i u_j+\partial_j u_i-\frac{2}{3}\delta_{ij}\partial_k u_k)+\zeta \delta_{ij}\partial_k u_k\, ,
\lb{20}
\ee
where $\rho$ is energy density, $p$ is pressure and $\eta$ and $\zeta$  are  the viscosity parameters.
Note that any term in $T_{\mu\nu}$ that is due to $g_{\mu\nu}$ or $\delta_{ij}$ will not contribute to $T^A_{\mu\nu}$  in (\ref{19}).
Taking  4-velocity in the form $u_\mu=(-1, u_i)$ one finds
\be
&&\hat{T}_{00}=\rho\, , \nonumber \\
&& \hat{T}_{0i}=-(\rho+p) u_i +\lambda [(\rho+p) \omega_i+\epsilon_{lki} u_ k\partial_l (\rho+p)]\nonumber \\
&&  \hat{T}_{ij}=T_{ij}+\lambda[-(\rho+p)\omega_i u_j+\epsilon_{kli}u_k\partial_l ((\rho+p) u_j)+   \eta \partial_i \omega_j+(i\leftrightarrow j)]\, ,
\lb{21}
\ee
where $\omega_i=\epsilon_{kli}\partial_k u_l$ is vorticity ($\vec{\omega}={\rm curl}\, \vec{u}$).

\bigskip

\noindent{\it  Modified Navier-Stokes equation as a Beltrami vector field.} For simplicity below we consider the non-relativistic fluid, so that in equations (\ref{20})-(\ref{21}) 
we use approximation $\rho+p\approx \rho$.
When the fluid is described by stress-energy tensor $T_{\mu\nu}$ (\ref{20}) the conservation law $\partial_0 T_{00}-\partial_j T_{j0}=0$ 
leads to continuity equation 
\be
\partial_t\rho=-\partial_j (\rho u_j)\, ,
\lb{22}
\ee
while the conservation law $\partial_0 T_{0i}-\partial_j T_{ij}=0$ is 
the Navier-Stokes equations for viscous fluid. Let us define vector
\be
  N_i\equiv\rho(\partial_t u_i+u^j \partial_j u_i)+\partial_i p-\partial_j \sigma_{ij}\, .
\lb{22-1}
\ee
Then for the classical stress tensor (\ref{20}) the Navier-Stokes equations are given by relation 
\be
N_i=0\, .
\lb{22-11}
\ee 
The consistency condition for the approximation under consideration requires that the modified stress tensor  $\hat{T}_{\mu\nu}$ (as defined in equation (\ref{19})) be conserved when considered on a Minkowski background metric. This may lead to certain modifications in the equations described above. In fact, equation (\ref{22}) remains unchanged for two reasons: (i) there is no modification in the $(00)$ component of the stress tensor, and (ii) although the $(0i)$ component is modified by $T^A_{0i}$  this modification is divergence-free, meaning  $\partial_j T^A_{0j}=0$, and thus does not alter equation (\ref{22}). On the other hand, the Navier-Stokes equations are now represented by the following relation:
\be
N_i-\lambda \epsilon_{kli}\partial_l N_k=0\, ,
\lb{22-3}
\ee
where $N_i$ is defined in (\ref{22-1}).  This equation is a modification of the classical equation (\ref{22-11}). As a byproduct  one has that $N_i$ is divergence free, $\partial_i N_i=0$. Equation (\ref{22-3}) has  a curious interpretation. Indeed, equation (\ref{22-3}) taken separately defines a Beltrami vector field $\vec{N}$:  $\vec{N}+\lambda \, \textup{curl}\vec{N}=0$. 
It serves as a source for the
ordinary Navier-Stokes equation (\ref{22-1}).

\bigskip

\noindent{\it Steady Beltrami velocity.} Consider incompressible fluid, $\rho={\rm const}$, so that $\partial_j u_j=0$. Then one has that $\partial_j \sigma_{ij}=\eta \Delta u_i$.
So that the modified Navier-Stokes equations  (\ref{22-1}) can be written in the form
\be
\partial_t \hat{u}_i+u_j\partial_j \hat{u}_i -\lambda \omega_j \partial_j u_i+\partial_i (\frac{p}{\rho})-\frac{\eta}{\rho}\Delta \hat{u}_i=0\, ,
\lb{23}
\ee
where $\hat{u}_i=u_i+\lambda \omega_i$, $\omega_i$ is the vorticity. There exists a steady, dissipation free, solution to this equation provided that $\hat{u}_i=0$. This defines the Beltrami velocity
\be
u_i-\lambda \epsilon_{kli}\partial_l u_k=0\, ,
\lb{24}
\ee
or, in vector form, $\vec{u}+\lambda \, {\rm curl}\,  \vec{u}=0$.  Equation (\ref{23}) then reduces\footnote{One here uses that  $u_j\partial_j u_i=\frac{1}{2}\partial_i u^2-(\vec{u}\times \textup{curl}\vec{u})_i$.} to
a  gradient  condition $\partial_i B=0$ of a function
\be
B=p+\frac{1}{2}\rho u^2\, 
\lb{25}
\ee
that  implies that  this function is constant in the fluid, $B=const$. This is a standard condition in the theory of Beltrami fields in fluid mechanics.
The solution to both equations (\ref{24}) and (\ref{25}) then describes a steady non-dissipative distribution of  fluid velocity.

\bigskip

\noindent{\it Generalized Gromeka-Beltrami flow and the Beltrami vorticity.}
As usual, one may re-express the Navier-Stokes equation (\ref{23}) in the form that includes only velocity \cite{LL}. For that one should simply apply the operator ${\rm curl}$ to
equation (\ref{23}). It is convenient to express the resulting equation in the vector form,
\be
&&\partial_t \hat{\vec{\omega}}-\nu \Delta  \hat{\vec{\omega}}-\vec{\Omega} -\lambda \, \textup{curl}\,  \vec{\Omega}=0\, , \nonumber \\
&&\hat{\vec{\omega}}=\vec{\omega}+\lambda \, \textup{curl}\, \vec{\omega}\, , \  \  \  \nonumber \\
&&\vec{\Omega}=\textup{curl}(\vec{u}\times \vec{\omega})\, , \ \ \nu=\frac{\eta}{\rho}\, .
\lb{26}
\ee
The Beltrami flow is usually defined by condition $\vec{u}\times \vec{\omega}=0$ that implies the mutual proportionality of velocity vector $\vec{u}$ and vorticity vector 
$\vec{\omega}$. Sometimes it is useful to  define the flow by a more general condition $\vec{\Omega}=0$, see \cite{NS}. 
The corresponding flow is the Gromeka-Beltrami flow. 
Looking at equation (\ref{26}) it seems rather natural to impose a yet more general  condition
$\vec{\Omega} +\lambda \, \textup{curl}\,  \vec{\Omega}=0$. Then  equation (\ref{26}) reduces to the heat equation
\be
\partial_t \hat{\vec{\omega}}-\nu \Delta  \hat{\vec{\omega}}=0\, .
\lb{26-1}
\ee
The solution to the heat equation  decays  over time, asymptotically  approaching
$\hat{\vec{\omega}}=0$. This indicates that the asymptotic vorticity  behaves as  a Beltrami vector field, 
\be
\vec{\omega}+\lambda \, \textup{curl}\, \vec{\omega}=0\, .
\lb{27}
\ee
This equation is consistent  with, hough not equivalent to,  equation (\ref{24}).

\section{Concluding remarks} 

The parameter $\Lambda$ (with dimensions of energy) that appears in the stress-energy tensor induced by the chiral anomaly (\ref{14-3}) is proportional to the chiral chemical potential $\mu_5$, similarly to the parameter $\kappa$ in the Chiral Magnetic Effect. The estimated value of the chiral chemical potential in measurable media is $\mu_5\simeq 500 \, {\rm Mev}$, which is reasonably low, making the momentum and stress fluxes induced by the chiral anomaly potentially observable in chiral media.
On the other hand, the parameter $\lambda$ (with dimensions of length), which arises in the current discussion of gravitational backreaction effects and Beltrami-type fields in the gravitational context, is related to $\mu_5$ as $\lambda \simeq G \mu_5$, similar to the relationship between gravitational radius and mass/energy (given by $\mu_5$). As a result, the numerical value of $\lambda$ in physical systems accessible on Earth is expected to be extremely small. This raises the question of whether the backreaction effects discussed here can be observed.

It should be noted, however, that in fluid mechanics, the streamlines of Beltrami vector fields form certain knotted structures. In fact, as shown in mathematical works (see \cite{knot}), any prescribed link in $\mathbb{ R}^3$  can be realized as a streamline of a Beltrami field. This property allows for speculation that even with microscopic values of $\lambda$, sufficiently macroscopic knotted field configurations could be observed experimentally, either in quark-gluon plasma or in Dirac and Weyl semimetals.
Additionally, Beltrami vector flow exhibits chaotic behavior, as evidenced by the ABC flow \cite{ABC}. This could be another important feature, potentially persisting in the geodesic flow of the gravitational configurations considered here. Such chaotic behavior could be detected by observing the geodesic trajectories of test particles.

The gravitational spheromak discussed in this note appears in the linearized version of the modified Einstein equations. It would be interesting to investigate whether this solution survives in the non-linear theory and, if so, whether it is stable. The mechanism for the formation of gravitational spheromaks remains an open question.

A potentially intriguing observation is that gravitational waves may serve as a source for the gravitational Beltrami tensor fields describing a spheromak. This suggests a possible mechanism for creating a spheromak as a gravitational wave passes through a chiral medium, such as the interior of a large astrophysical object. These and other potential applications of the findings presented here warrant further study.

\newpage
\appendix
\section{Modified Einstein equations in weak field \\  approximation}
\setcounter{equation}0
\numberwithin{equation}{section}
Here we follow the analysis present in \cite{SC}. We use the following parametrization for the  metric perturbations $h_{\mu\nu}$: $h_{00}=-2\Phi$, $h_{0i}=w_i$, $h_{ij}=2s_{ij}-2\delta_{ij} \Psi$, where $s_{ij}\delta^{ij}=0$.
One can use the coordinate freedom and impose the conditions: $\partial_i w_i=0$ and $\partial_j s_{ij}=0$.  So that $s_{ij}$ represents the transverse-traceless component of the metric.The modified Einstein equations for the metric perturbations is 
\be
G_{\mu\nu}=2\lambda\epsilon_{0\beta\sigma (\mu}\partial^\sigma R_{\nu )}^{\ \beta}\, ,
\lb{a1}
\ee
where $G_{\mu\nu}=R_{\mu\nu}-\frac{1}{2}g_{\mu\nu}R$ is the Einstein tensor.
In components it takes the form
\be
&&G_{00}=0\, , \nonumber\\
&&G_{0i}={\lambda}\epsilon_{kli}\partial_l R_{0k}\, , \nonumber\\
&&G_{ij}={\lambda}(\epsilon_{kli}\partial_l R_{kj}+(i \leftrightarrow  j))\, .
\lb{aa2}
\ee
In the weak field approximation one has that
\be
&&G_{00}=2\Delta \Psi\, , \nonumber \\
&&G_{0i}=-\frac{1}{2}\Delta w_i+2\partial_{0}\partial_i\Psi \, , \nonumber \\
&&G_{ij}=(\delta_{ij}\Delta- \partial_i \partial_j)(\Phi-\Psi)-\partial_0\partial_{(i}w_{j)}+2\delta_{ij}\partial_0^2\Psi -\Box s_{ij}\, ,
\lb{a3}
\ee
where $\Delta=\partial_i \partial_i$ and $\Box=-\partial_0^2+\Delta$.
For the Ricci tensor one has
\be
&&R_{00}=\Delta \Phi+3\partial_0^2\Psi\, , \nonumber \\
&&R_{0i}=-\frac{1}{2}\Delta w_i+2\partial_0 \partial_i \Psi\, , \nonumber \\
&&R_{ij}=-\partial_i\partial_j(\Phi-\Psi)-\partial_0\partial_{(i}w_{j)}+\delta_{ij}\Box \Psi -\Box s_{ij}\, .
\lb{a4}
\ee
Equations (\ref{a1})-(\ref{aa2}) then take the form
\be
&&(00): \  \  \Delta \Psi=0\, , \nonumber \\
&&(0i): \  \  (-\frac{1}{2}\Delta w_i+2\partial_0\partial_i\Psi)={\lambda}\epsilon_{kli}\partial_l(-\frac{1}{2}\Delta w_k+2\partial_0\partial_k\Psi)\, , \nonumber \\
&&(ij): \  \  (\delta_{ij}\Delta -\partial_i \partial_j)(\Phi-\Psi)-\partial_0\partial_{(i}w_{j)}+2\delta_{ij}\partial_0^2\Psi -\Box s_{ij}\nonumber \\
&&=-{\lambda}[\epsilon_{kli}(\partial_0\partial_l\partial_{(k}w_{j)}+\Box\partial_l s_{kj})+(i \leftrightarrow  j)]\, .
\lb{a5}
\ee
Solution of the Laplace equation $\Delta\Psi=0$ does not have non-trivial bounded solutions in $R^3$. So that one has that $\Psi=0$. Then taking the trace of the last equation in (\ref{a5})
one finds that $\Delta\Phi=0$. This implies that  $\Phi=0$. The second equation in (\ref{a5}) can be solved as follows,
\be
w_i-{\lambda}\epsilon_{kli}\partial_l w_k=w^0_i\, , \ \ \ \Delta w^0_i=0\, .
\lb{a6}
\ee
One  thus has that $w^0_i=0$ and, hence, $w_i$ is a Beltrami vector field,
\be
w_i={\lambda}\epsilon_{kli}\partial_l w_k\, .
\lb{a7}
\ee
Taking into account (\ref{a7}) the last equation in (\ref{a5}) takes the form (terms with $w_i$ are mutually cancelled  in (\ref{a5}) due to (\ref{a7})),
\be
\Box (s_{ij}-{\lambda}(\epsilon_{kli}\partial_l s_{kj}+(i \leftrightarrow  j)))=0\, .
\lb{a8}
\ee
This equation can be solved as
 follows
 \be
 s_{ij}-{\lambda}(\epsilon_{kli}\partial_l s_{kj}+(i \leftrightarrow  j))=s^0_{ij}\, , \  \  \  \Box s^0_{ij}=0\, .
 \lb{a9}
 \ee
 Tensor $s^0_{ij}$ thus represents the usual propagating gravitational wave with two polarizations. In cartesian coordinates the solutions for $s^0_{ij}$ are plane waves. So that components of
 $s^0_{ij}$ are bounded (although not integrable).  Assuming that $s^0_{ij}=0$ (no gravitational wave is present) one has that
 \be
  s_{ij}-{\lambda}(\epsilon_{kli}\partial_l s_{kj}+(i \leftrightarrow  j))=0\, .
  \lb{a10}
  \ee
  Since $s_{ij}\delta^{ij}=0$ and $\partial_j s_{ij}=0$ then this equation can be represented as
  \be
    s_{ij}-2\lambda\epsilon_{kli}\partial_l s_{kj}=0\, .
    \lb{a11}
    \ee
    The components of $w_i$ satisfy the Helmholtz equation $(\Delta +\frac{1}{\lambda^2})w_i=0$ and components of $s_{ij}$ satisfy equation $(\Delta +\frac{1}{4\lambda^2})s_{ij}=0$.
    Solutions to the Helmholtz equations are bounded  in $\mathbb{R}^3$ and, hence, represent the eligible physical degrees of freedom. In this aspect they are similar to the solutions to the
    wave equation.

\section{Solutions to Beltrami vector equations}
\setcounter{equation}0
\numberwithin{equation}{section}
Consider the Beltrami vector equation,
\be
v_i=\lambda \epsilon_{kli}\partial_l v_k\, ,
\lb{b1}
\ee
and represent solution to this equation that is changing in direction $x_1$ as 
\be
v^{(1)}_i=H_i^+ e^{\frac{ix_1}{\lambda}}+H^-_i e^{-\frac{ix_1}{\lambda}}\, ,
\lb{b2}
\ee
where $H^\pm_i$ are some  (complex) constants. 
Substituting this ansatz into equation (\ref{b1}) one finds
\be
H_1^\pm=0\, , \ \  H^\pm_2=\pm i H^\pm _3\, .
\lb{b3}
\ee
The reality condition for the vector components $v_i^{(1)}$ implies that $\overline{H^-_i}=H^+_i$. So that the only independent components 
are those of $H^+_3=\frac{1}{2}(\alpha+i\beta )$ and we find that
\be
&&v^{(1)}_1=0\, , \nonumber \\    
&&v^{(1)}_2=-\alpha \sin\frac{x_1}{\lambda}-\beta \cos \frac{x_1}{\lambda}\, , \nonumber \\
&&v_3^{(1)}=\alpha \cos \frac{x_1}{\lambda}-\beta\sin\frac{x_1}{\lambda}\, .
\lb{b4}
\ee
Repeating same procedure for vector field components changing in direction of $x_2$ and in direction $x_3$ we find
\be
&&v^{(2)}_1=\gamma \sin\frac{x_2}{\lambda}+\epsilon \cos \frac{x_2}{\lambda}\, , \nonumber \\    
&&v^{(2)}_2=0,  \nonumber \\
&&v_3^{(2)}=\gamma \cos \frac{x_2}{\lambda}-\epsilon\sin\frac{x_2}{\lambda}\, .
\lb{b44}
\ee
and
\be
&&v_1^{(3)}=-\chi \sin\frac{x_3}{\lambda}-\eta\cos\frac{x_3}{\lambda} \, , \nonumber \\
&&v_2^{(3)}=\chi \cos\frac{x_3}{\lambda}-\eta\sin \frac{x_3}{\lambda} \, , \nonumber \\
&&v_3^{(3)}=0\, .
\lb{b5}
\ee
The solution to linear equation (\ref{b1}) can be a combination of all three fields,
$v_i=v_i^{(1)}+v_i^{(2)}+v_i^{(3)}$,
\be
&&v_1=\gamma \sin\frac{x_2}{\lambda}+\epsilon \cos \frac{x_2}{\lambda}-\chi \sin\frac{x_3}{\lambda}-\eta\cos\frac{x_3}{\lambda} \, , \nonumber \\
&&v_2=-\alpha \sin\frac{x_1}{\lambda}-\beta \cos \frac{x_1}{\lambda}+\chi \cos\frac{x_3}{\lambda}-\eta\sin \frac{x_3}{\lambda} \, , \nonumber \\
&&v_3=\alpha \cos \frac{x_1}{\lambda}-\beta\sin\frac{x_1}{\lambda}+\gamma \cos \frac{x_2}{\lambda}-\epsilon\sin\frac{x_2}{\lambda}\, .
\lb{b6}
\ee
This solution is determined by six parameters: $\alpha, \, \beta, \, \gamma, \, \epsilon, \, \chi, \, \eta$.

For $\lambda=-1$, imposing extra conditions $\gamma=\eta=\beta=0$, solution (\ref{b6}) reduces to the ABC (Arnold-Beltrami-Childress)  vector fields widely discussed in the literature,
\be
&&v_1=A\sin x_3+C\cos x_2\, , \nonumber \\
&&v_2=B\sin x_1 +A \cos x_3\, , \nonumber \\
&&v_3=C\sin x_2 +B\cos x_1\, ,
\lb{c0}
\ee
where $A=\chi$, $B=\alpha$ and $C=\epsilon$ are constants.  One finds that $\vec{v}=\textup{curl}\, \vec{v}$ and $\Delta \vec{u}+\vec{u}=0$.
The ABC flow is then given by equation $\frac{d \vec{x}}{d\tau}=\vec{v}(x)$, $\vec{x}\equiv \{ x_1, x_2, x_3 \}$.

\section{Solutions to Beltrami tensor equations}
\setcounter{equation}0
\numberwithin{equation}{section}
Consider now the tensor Beltrami equation,
\be
h_{ij}=2{\lambda}\epsilon_{kli}\partial_l  h_{kj}\, .
\lb{c1}
\ee
The index $j$ in this equation is free and thus the tensorial equation (\ref{c1}) is formally equivalent to three vector Beltrami equations plus the condition of symmetry $h_{ij}=h_{ji}$.
We use the solution (\ref{b6}) for Beltrami vector fields (where we have to replace $\lambda$ by $2\lambda$) found in appendix B and simply add index $j$,
\be
&&h_{1j}=\gamma_j \sin\frac{x_2}{2\lambda}+\epsilon_j \cos \frac{x_2}{2\lambda}-\chi_j \sin\frac{x_3}{2\lambda}-\eta_j\cos\frac{x_3}{2\lambda} \, , \nonumber \\
&&h_{2j}=-\alpha_j \sin\frac{x_1}{2\lambda}-\beta_j \cos \frac{x_1}{2\lambda}+\chi_j \cos\frac{x_3}{2\lambda}-\eta_j\sin \frac{x_3}{2\lambda} \, , \nonumber \\
&&h_{3j}=\alpha_j \cos \frac{x_1}{2\lambda}-\beta_j\sin\frac{x_1}{2\lambda}+\gamma_j \cos \frac{x_2}{2\lambda}-\epsilon_j\sin\frac{x_2}{2\lambda}\, .
\lb{c6}
\ee
The symmetry conditions $h_{12}=h_{21}$, $h_{13}=h_{31}$ and $h_{23}=h_{32}$ give us the constraints,
\be
&&\gamma_2=0\, , \ \epsilon_2=0\, , \  \chi_3=0\, , \ \eta_3=0\, , \ \alpha_1=0\, , \ \beta_1=0\, , \nonumber \\
&&\gamma_1=\epsilon_3\, , \ \gamma_3=-\epsilon_1\, , \ \chi_2=\eta_1\, , \ \eta_2=-\chi_1\, , \ \alpha_2=-\beta_3\, , \ \alpha_3=\beta_2\, ,
\lb{c7}
\ee
Thus, there remains six independent parameters: $\beta_2, \ \beta_3, \ \epsilon_1, \ \epsilon_3, \ \chi_1, \ \eta_1$. The  components of the  Beltrami tensor field then read,
\be
&&h_{12}=C\cos\frac{x_3}{2\lambda}-D\sin\frac{x_3}{2\lambda}=h_{21}\, , \nonumber \\
&&h_{13}=-B\sin\frac{x_2}{2\lambda}+A\cos \frac{x_2}{2\lambda}=h_{31}\, , \nonumber \\
&&h_{23}= -E\sin\frac{x_1}{2\lambda}-F\cos \frac{x_1}{2\lambda}=h_{32}\, , \nonumber \\
&&h_{11}=A\sin \frac{x_2}{2\lambda}+ B \cos \frac{x_2}{2\lambda}-C \sin\frac{x_3}{2\lambda}-D \cos\frac{x_3}{2\lambda}\, , \nonumber \\
&&h_{22}=F\sin \frac{x_1}{2\lambda}-E \cos\frac{x_1}{2\lambda}+D\cos\frac{x_3}{2\lambda}+C\sin\frac{x_3}{2\lambda}\, , \nonumber \\
&&h_{33}=E\cos\frac{x_1}{2\lambda}-F\sin\frac{x_1}{2\lambda}-B\cos \frac{x_2}{2\lambda}-C\sin\frac{x_2}{2\lambda}\, ,
\lb{c8}
\ee
where $A=\epsilon_3$, $B=\epsilon_1$, $C=\chi_1$, $D=\eta_1$, $E=\beta_3$, $F=\beta_2$ are six parameters that determine the solution.
We note that $h_{11}+h_{22}+h_{33}=0$ so that $h_{ij}$ (\ref{c8})  is automatically traceless as expected.

\end{document}